\documentclass[letterpaper,12pt,oneside,onecolumn]{article}%
\usepackage{geometry}%
\usepackage{titlesec}%
\titleformat{\section}[hang]{\normalfont\normalsize\bfseries}{\thesection}{12pt}{\centering}%
\titleformat{\subsection}[display]{\normalfont\normalsize}{\thesubsection}{12pt}{\underline}%
\titleformat{\subsubsection}[runin]{\normalfont\normalsize}{\thesubsubsection}{12pt}{\underline}%
\newcommand{\PaperTitle}[1]{%
\begin{center}%
    \begin{large}%
        \textbf {#1} \\%
    \end{large}%
\end{center}%
}%
\newcommand{\AuthorList}[1]{%
\begin{center}%
    {#1} \\%
\end{center}%
}%
\newcommand{\AuthorAffiliations}[1]{%
\begin{center}%
    {#1} \\%
\end{center}%
}%
\newcommand{\Keywords}[1]{%
\begin{center}%
   Keywords: {#1} \\%
\end{center}%
}%
\usepackage{graphics}
\usepackage{epsfig}
\begin{document}%
\PaperTitle{Fermi surface effects in solid-solid phase transitions in simple metals under pressure
}%
\AuthorList{G.J.Ackland}%
\AuthorAffiliations{School of Physics and Centre for Science at
Extreme
Conditions, The University of Edinburgh, Mayfield Road, Edinburgh, EH9 3JZ, UK}%
\Keywords{Pressure, cesium, lithium}%
\section{Abstract}
We discuss the behaviour of Group 1 and 2 elements under pressure,
describing crystallography based on diamond anvil pressure cell experiments
and electronic
structure calculations.  In addition to these precise methods, we
discuss more intuitive pictures of both the electronic driving force
for the transitions, and the structural motifs which optimise the
electronic energy by opening pseudogaps at the Fermi energy.
\section{Introduction} 
Phase transitions occur when the relative stability of two putative phases
changes.  The stable phase is the one with lowest Gibbs free energy
(G), and in most familiar temperature-driven cases the transitions
arises dues to the relative importance of energy U and entropy S, 
$G=U-TS$. 

Transitions may also be caused by external strain and magnetic field,
however because of the ability of a crystal to support a shear, or to
screen a magnetic field, such transitions lead to formation of domain
microstructure, and the transformation is dominated by interface
effects between parent and daughter phase, which may co-exist.  Such
transitions are still described by the Gibbs free energy, with
additional terms describing the strain and/or magnetic energies, which
compete with the internal energy and entropy:
$ 
G=U-TS+\sigma\epsilon+\mu B$

Now additional complications arise from the fact that the strain field
is a tensor, and the magnetic field is a vector.

Under isotropic pressure,  the relative importance of various contributions to
cohesion varies continuously.  It is therefore possible to find phase
transitions between structures with different electronic structures,
and for these transitions to be induced by an externally applied field
(the pressure) rather than arising from phase separation.  Because the
phase separation necessarily involves diffusion, the possibility that
a diffusionless transition between crystal phases may be possible is
suppressed.  Thus pressure transitions give us a way to study
structural transitions driven by changes in the electronic structure,
without the complicating fact of diffusion.

Experimental pressures in the GPa range can routinely be obtained,
thus the free energy contribution from the PV term can be much larger
than that arising from TS.  As a consequence, temperature effects are
routinely ignored when considering high-P phase transitions, and we minimise the enthalpy 
$G\approx H=U+PV$.

Extensive experimental\cite{Li,Na,Cs,Rb,K,Sr,Ba} and theoretical studies\cite{NT99,CN01,skr,irm,Ahuja-CsRb} on the alkali and
alkaline earth elements reveals a similar theme to the sequence of 
transitions with
pressure.  Typically the materials have a low-pressure, close-packed
structure (either fcc or bcc - both have very similar energies), then
a range of less-densely packed structures at intermediate pressure,
and finally a return to close packing at the highest pressures.  This
apparently counter-intuitive behaviour, with phases going from
close-packed to open and back to close packed, is also associated with
low-T superconductivity\cite{LiCsGaRb-supercond} and a drop in the
conductivity in the intermediate regime\cite{resist}.  It can be
intuitively understood as follows: the low temperature "close packed"
phases actually represent atoms being as far apart as possible,
minimising their electrostatic repulsion in a delocalised
free-electron gas. The atoms have relatively high entropy, manifested
as very low phonon frequencies and elastic moduli.  As pressure
increases, the stronger interactions between atoms and electrons
favours structures which maximise this effect, in particular those
which open a pseudogap at the Fermi energy.  At still higher pressures
the Pauli repulsion between the core electrons becomes dominant, and
again close packed structures are obtained. Thus the interesting
region crystallographically lies at ``intermediate'' pressure.

 We use the term ``intermediate'' because the actual 
pressures involved vary from element to element, i.e. the
absolute GPa value is system dependent
with lighter elements having larger absolute values.

The band structure stabilisation of certain crystal structures by
interactions between the Fermi surface and Brillouin zone has been
known for many years \cite{MottJones} is familiar in many metallic
compounds, from Hume-Rothery phases to quasicrystals, at special
values of the number of valence electrons.  It is only one of 
the competing effects, and the advent of total energy calculation has 
blurred the distinction between them, however,  the 
Kohn-Sham Hamiltonian\cite{ks}  comprises a
band structure term (the eigenvalues of non-interacting electrons
moving in an effective field), an Ewald sum, 
exchange correlation and Hartree energies.  The dependence on density
($\rho$) is as follows: $\rho^{1/3}$ for the coulombic terms
(Hartree, ion-ion, ion-electron), $\rho^0$ for the exchange
correlation (neglecting ``non-linear core corrections''),
$\rho^{2/3}$ for kinetic energy.  Aside from the exclusion force 
from orthogonalisation, the only energy term which increases 
linearly with density is that due to energy splitting
from interaction between plane waves and the lattice at the Fermi level.
$\Delta E = \pm \int_\Omega e^{i{\bf k.r}} V({\bf q})  e^{i{\bf k'.r}} d^3{\bf 
r}$

For group 1 and 2 elements the relative importance of this effect
varies continuously with application of pressure. Although one cannot
easily continuously vary the valence electron density in a sample
chemically, pressure provides an external field enabling us to find
and traverse the region where Fermi surface effects dominate.

\section{How elements maximise Fermi surface Brillouin zone interactions}

Having established that the total energy can be reduced by finding
structures which maximise Fermi-surface effects, there remains the
problem of determining exactly what kind of structure arises.  There
are close parallels with the powder crystallography experimental
situation here - we know theoretically which scalar Bragg vectors
would be favoured ($2k_F$), and the problem is to reconstruct a crystal
which optimise these.  While some structures can be guessed as
distortions of simple structure, it turns out that nature has some
admirably complex structures which are solutions to the FSBZ
optimisation problem.

\subsection{Distortions of simple structures}

\begin{figure}[ht]
\protect{\includegraphics[width=\columnwidth]{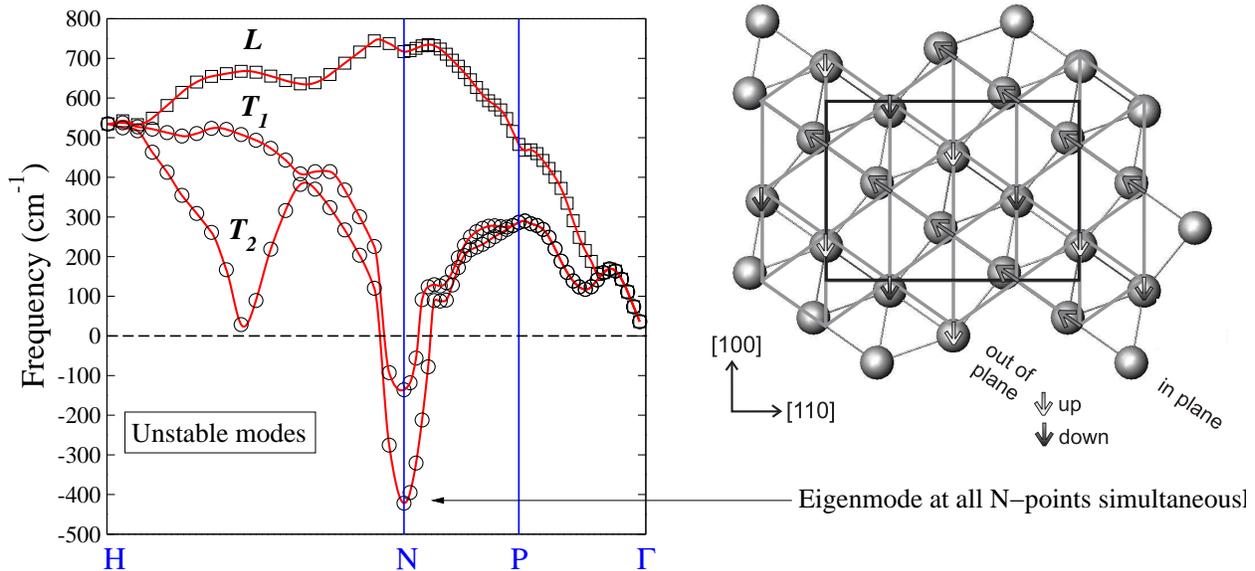}}
\caption{
Calculated phonon spectrum for bcc Li at 40GPa.  The ``negative''
frequency phonons correspond to unstable modes, i.e. distortions which
lower the energy of the crystal.
\protect\label{fig:Li}}
\end{figure}

The bcc structure provides a close packed reciprocal lattice, and is
therefore a good starting point to hunt for FSBZ-optimising
structures.  Lattice dynamics calculations\cite{phon} enable us to determine
mechanical instabilities of bcc, or other high-symmetry structures.
Lithium provides a good example\cite{irm} - Figure.\ref{fig:Li} shows a
phonon spectrum calculation for bcc lithium under pressure, showing
that the structure is unstable, and the stable structure associated
with allowing this instability to relax. These calculations are done using density functional theory and the plane wave pseudopotential method PW-DFT\cite{payne} 
with phonons calculated by finite displacements\cite{phon}, full details are given elsewhere\cite{irm}

 The distorted structure turns out to be the
cI16 crystal structure observed in lithium\cite{Li} at intermediate pressures.
The energy gained in the distortion comes from opening a pseudogap at
the Fermi level.  The mechanism by which such a structure can be
reached from bcc is straightforward.

\subsection{Spontaneous interstitial formation}

\begin{figure}[ht]
\protect{\includegraphics[width=\columnwidth]{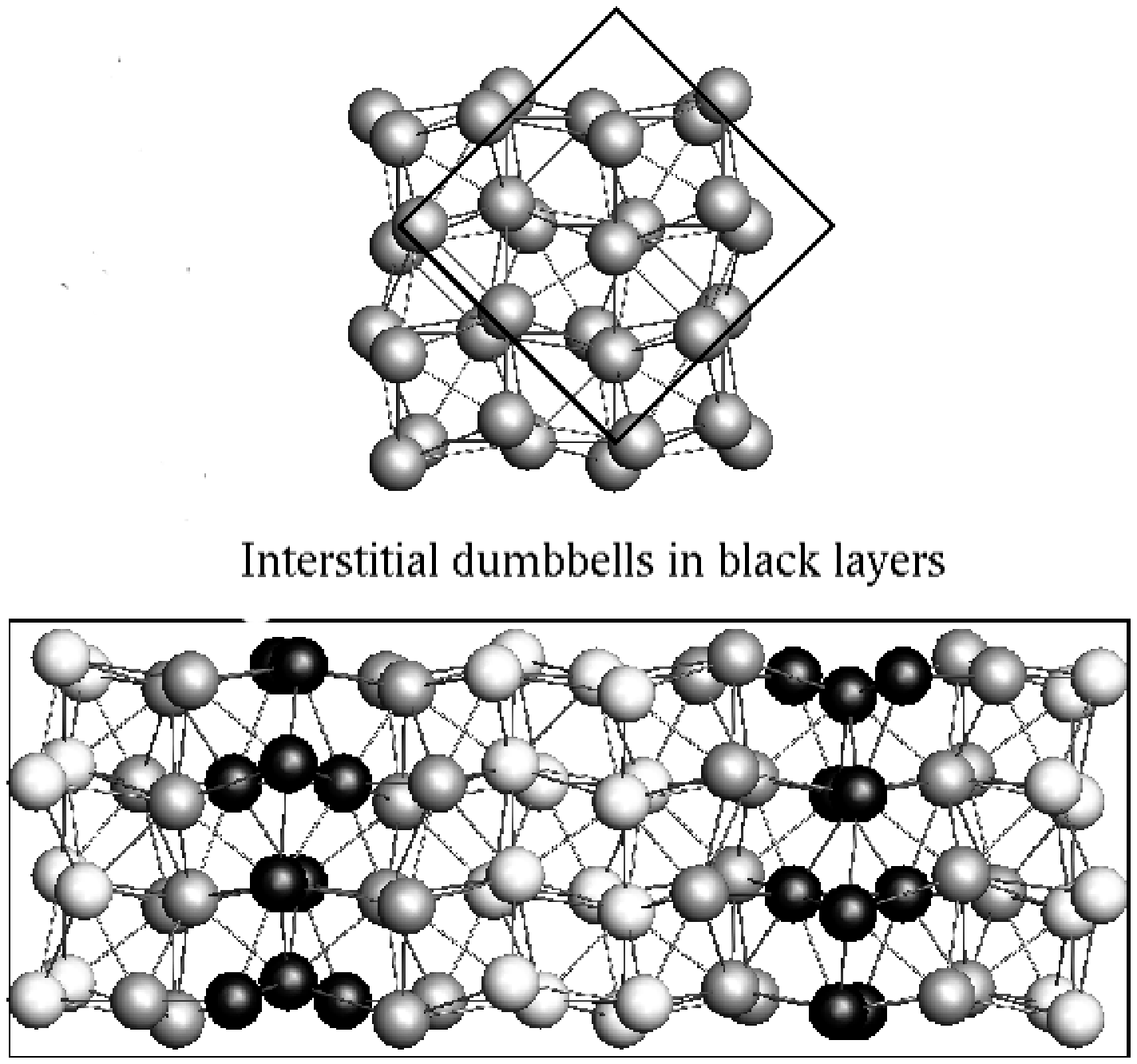}}
\caption{
Relationship between intermediate pressure crystal structures in Li (top, with black diamond showing doubled, distorted bcc cell))
and Cs bottom. Four interstitial dumbbells (two atoms on one site, pointing along the longer axis) have been inserted in the black layers,
 Rubidium shows similar behaviour.
\protect\label{fig:LiCs}}
\end{figure}

The distorted bcc structure of Li may provide the option of
multiple FSBZ interactions.  However, it may prove impossible to match
the necessary crystal density with the required number of atoms.
Nature's ingenious solution here is to incorporate interstitial atoms
onto the distorted bcc structure.  Figure.\ref{fig:LiCs} 
shows the relationship
between the intermediate crystal structure of lithium and that of
caesium, found both experimentally\cite{Cs} and theoretically\cite{Ahuja-CsRb}
and confirmed by our own PW-DFT calculation.
 This shows that the Cs structure can be thought of as
being created from the distortion of the bcc structure into the cI16
structure, combined with the formation of planes of interstitials
(black layers).  No work has been done on the mechanism for this
transformation, in particular whether the distortion can occur in a
diffusionless manner while the interstitial array takes longer due to  
being diffusive. 

\subsection{Incommensurate structures}

Another way to increase FSBZ interactions is to increase the
dimensionality of the reciprocal space.  This is observed when an
incommensurate structure (sometimes called a charge density wave) is
formed.  This conventional type of structure is a modulation of a
simple crystal structure, as observed in high-pressure selenium.
Simple elemental metals have a far more complex modulation: the
``hotel'' structure. Some ``host'' atoms form a zeolite-like crystal
structure, with sizable, usually linear, pores. The remaining atoms
then form ``guest'' chains inside the pores. Such structures can then
be described by two inter-penetrating lattices, incommensurate in one
axis, which then require a four dimensional reciprocal space to
describe.  By transferring atoms between ``guest'' and ``host''
structures, the shape of the 4D reciprocal lattice can be moulded to
fit the Fermi surface, and thus lower the energy.

\begin{figure}[ht]
\protect{\includegraphics[width=\columnwidth]{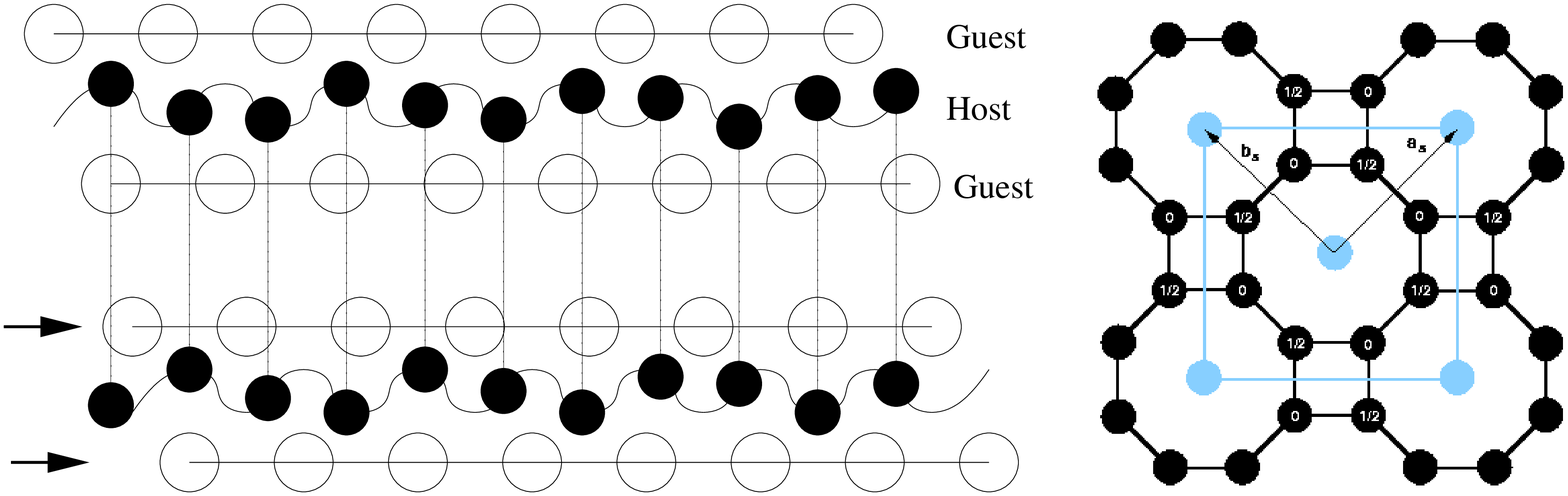}}
\caption{Schematic incommensurate crystal structure in barium IV, 
showing inter-penetrating ``guest'' and ``host'' structures (a) shows chains end-on fitting into the host structure (b) shows incommensurate guests and hosts along the c-axis.  The guest chains are regularly arranged with respect to each other, but incommensurate with the hosts.  Rubidium, potassium and sodium show similar guest-host behaviour.
\protect\label{BaIV}}
\end{figure}

\subsection{What about s-d transfer?}
The traditional picture of the complex intermediate phases is one in
which electrons transfer from ``delocalised, large, spherical''
s-orbitals to ``localised, compact, angular'' d-orbitals.  It is
important to realise that both this and the FSBZ model are just
pictures - one can of course make projections of the electronic
wavefunctions onto localised orbitals and find that as pressure
increases there is a transfer from s to d.  Historically, the key
experimental ``proof'' of this was the report isostructural fcc-fcc
transition in Cs, which of course cannot introduce extra FSBZ effects.
However, this is undermined by the recent experimental determination
that there is no such transition - the intermediate phase 
was simply misidentified.  In
favour of the FSBZ picture are:
\begin{itemize}
\item that the energetics can be
described by density functional theory (which is known to fail for
strongly localised electrons)

\item the similarity between the structures
in high pressure Li and those in Cs and Rb (despite the absence of a
2d electron shell!)

\item the drop in conductivity in the intermediate
phases associated with opening of the pseudogap

\item the experimental observation of structures with multiple strong diffraction peaks around $2k_F$.

\item The calculation of dips in the density of states at the Fermi level for complex structures.
\end{itemize}

\subsection{Consequences for transition mechanisms}

The phase transition mechanisms between complex phases of the elements
cannot be done by simple displacements in all cases, however in simple
metals the metallic binding and low melting point/shear moduli mean
that the transitions do occur on the timescale of at most few hours
after pressurization.  Since no composition change is required for an
element, relatively small atomic motions could effect the transitions,
however to my knowledge there has been no work in this area, either
experimentally or theoretically.

It is interesting to note that the FSBZ energy reduction
comes from long range periodic interactions of delocalised distortion 
and delocalised electrons.  It is therefore unclear whether this would 
be approached by a localised nucleation and growth process, as envisaged 
in most continuum mechanics approaches, or whether the nucleation should be 
thought of in reciprocal space, and the ``growth'' as the freezing in of a 
single in-phase long range mode.

\section{Conclusions}
We have shown that from scaling relations one might expect a pressure
regime in which Fermi Surface-Brillouin zone interactions dominate the
energy differences between crystal structures of similar density.  We
have then shown how experimental and density-functional theory data
from our own work and the literature support this picture for the
complex intermediate phases of the group I and II elements.  The mechanism by which these phase transformations occur is as yet unknown.

{\bf Acknowledgements:}  The author would like to thank 
C.Henjy, M.I.MacMahon, S.Falconi, O.Degytereva, S.K.Reed, I.R.MacLeod and V.Degytereva for useful discussions.

\end{document}